\documentclass[11pt,twoside]{article}
\usepackage[pdftex]{graphicx}
\usepackage{amsmath}
\usepackage{amssymb}
\usepackage{cite}

\begin{document}
\newcommand{\pst}{\hspace*{1.5em}}

\newcommand{\rigmark}{\em Journal of Russian Laser Research}
\newcommand{\lemark}{\em Volume 30, Number 5, 2009}

\newcommand{\be}{\begin{equation}}
\newcommand{\ee}{\end{equation}}
\newcommand{\bm}{\boldmath}
\newcommand{\ds}{\displaystyle}
\newcommand{\bea}{\begin{eqnarray}}
\newcommand{\eea}{\end{eqnarray}}
\newcommand{\ba}{\begin{array}}
\newcommand{\ea}{\end{array}}
\newcommand{\arcsinh}{\mathop{\rm arcsinh}\nolimits}
\newcommand{\arctanh}{\mathop{\rm arctanh}\nolimits}
\newcommand{\bc}{\begin{center}}
\newcommand{\ec}{\end{center}}

\thispagestyle{plain}

\label{sh}


\begin{center} {\Large \bf
\begin{tabular}{c}
SCHR\"{O}DINGER PICTURE ANALYSIS OF THE BEAM SPLITTER: 
\\[-1mm]
AN APPLICATION OF THE JANSZKY REPRESENTATION
\end{tabular}
 } \end{center}

\bigskip

\bigskip

\begin{center} {\bf
Stephen M. Barnett}\end{center}

\medskip

\begin{center}
{\it
School of Physics and Astronomy, University of Glasgow, Glasgow G12 8QQ, U.K.}
\smallskip

Corresponding author e-mail:~~~stephen.barnett@glasgow.ac.uk\\
\end{center}

\begin{abstract}\noindent
The Janszky representation constructs quantum states of a field mode as a superposition of coherent states on a 
line in the complex plane.  We show that this provides a natural Schr\"{o}dinger picture description of the interference between
a pair of modes at a beam splitter.
\end{abstract}

\medskip


\section{Introduction}
\pst
Surely there is no more ubiquitous device in quantum optics than the humble beam splitter.  A visitor to any laboratory
will doubtless see a large number of these arranged on an optical table, either standing on optical mounts or as coupled
fibres.  Many transformative experiments have relied on this device: three examples are the demonstration of single-photon
interference juxtaposed with the fact that the photon does not split to go both ways \cite{Alain}, the demonstration of squeezing using 
balanced homodyne detection \cite{Slusher,Shelby,Wu,LoudonKnight,Bachor} and the Hong-Ou-Mandel interference between a pair 
of photons \cite{HOM,LoudonJOSAB}.

The theory of the beam splitter is usually presented as a scattering of the annihilation and creation operators for the participating
field modes in which the operators for the outgoing modes are written in terms of those for the ingoing modes 
\cite{RodneyOC,RodneyBook,Methods}.  This is, in essence,
a Heisenberg picture theory, although it does provide a link between the input and output states by writing these in terms of the
action of creation operators on the vacuum state.

Among all the states of the radiation field, the coherent states are special 
\cite{RodneyBook,Methods,GlauberLett,SudarshanLett,GlauberCoh1,GlauberCoh2,Klauder,Nussenzveig,GlauberBook}: 
they provide the closest approximation to classical
fields, are minimum uncertainty states of the field quadratures and, of particular significance for us here, have a simple behaviour 
when combining at a beam splitter.  This simple behaviour makes coherent states the natural choice of basis for describing the
action of a beam splitter if we can describe the incoming and outgoing modes in terms of coherent states.

J\'{o}zsef Janszky and his colleagues provided an ingenious representation of some important quantum states of light formed as
line-integral superpositions of coherent states in the complex plane \cite{JJ90,JJ93,JJ94,JJ95}.  We apply this representation to 
provide a purely Schr\"{o}dinger
picture description of quantum interference, one in which the field annihilation and creation operators do not appear.

\section{Conventional theory of the beam splitter}
\begin{figure}[htbp] 
\centering
\includegraphics[width=12cm]{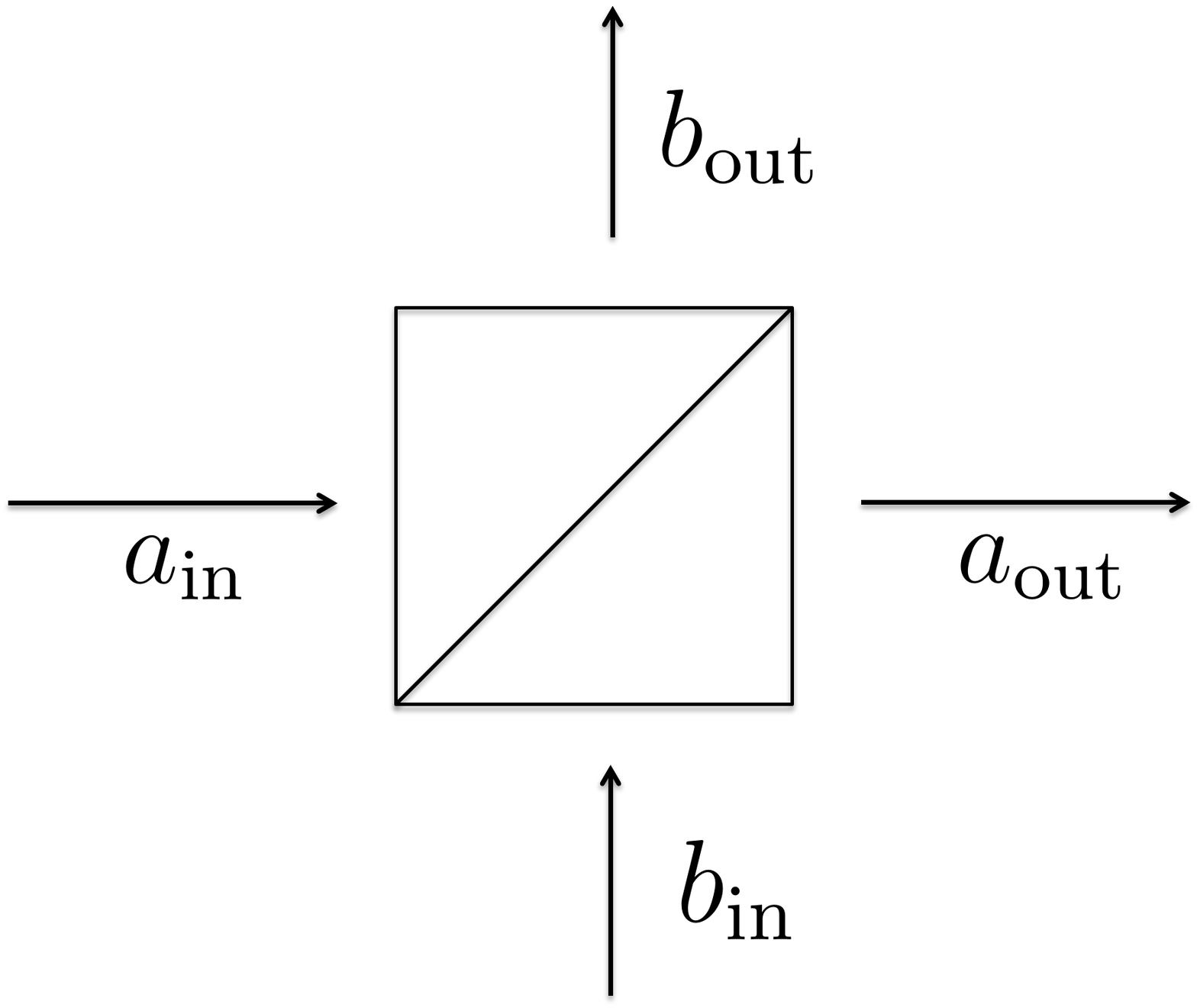}
\caption{Schematic of the interfering modes at a beam splitter.  The two input modes, $a_{\rm in}$ and $b_{\rm in}$
are superposed to form the two output modes, $a_{\rm out}$ and $b_{\rm out}$.} 
\label{fig:figure1}
\end{figure}

\pst
To provide some background and also a check on our later results, we begin with a short summary of the well-established
theory of the beam splitter as it is usually presented \cite{RodneyOC,RodneyBook}.  The device combines two overlapping 
input modes, $a_{\rm in}$ and
$b_{\rm in}$, to produce two output modes, $a_{\rm out}$ and $b_{\rm out}$, as depicted in Fig. \ref{fig:figure1}.
For simplicity we consider only a symmetric beam splitter for which the output annihilation operators are related to those of
the input modes by 
\begin{equation}
\label{Eqn1}
\left(\begin{array}{c}
\hat{a}_{\rm out} \\
\hat{b}_{\rm out}\end{array}\right)  =
\left(\begin{array}{cc}
t & r \\
r & t \end{array}\right) 
\left(\begin{array}{c}
\hat{a}_{\rm in} \\
\hat{b}_{\rm in}\end{array}\right) .
\end{equation}
The transmission and reflection coefficients, $t$ and $r$, are not independent, being restricted by the constraints of unitarity 
so that
\begin{eqnarray}
\label{Eqn2}
|t|^2 + |r|^2 &=& 1 \nonumber \\
tr^* + rt^* &=& 0 .
\end{eqnarray}
It is interesting to note that these conditions may be relaxed if the device includes losses, but we consider here only an
ideal and lossless beam splitter.  The transformation (\ref{Eqn1}) is readily inverted:
\begin{equation}
\label{Eqn3}
\left(\begin{array}{c}
\hat{a}_{\rm in} \\
\hat{b}_{\rm in}\end{array}\right)  =
\left(\begin{array}{cc}
t^* & r^* \\
r^* & t^* \end{array}\right) 
\left(\begin{array}{c}
\hat{a}_{\rm out} \\
\hat{b}_{\rm out}\end{array}\right) 
\end{equation}
and we can use this relationship, together with the conjugate relationship for the creation operators to write 
our output state in terms of the input state.  The principle is best explained by some simple examples.

\subsection{Interfering coherent states}

The coherent states, $|\alpha\rangle$, have played a major role in the development of quantum optics, starting with
their introduction by Glauber in the quantum theory of optical coherence \cite{RodneyBook,Methods,GlauberLett,SudarshanLett,GlauberCoh1,GlauberCoh2,Klauder,Nussenzveig,GlauberBook}.  
The coherent states are right-eigenstates 
of the annihilation operator and are characterised by this complex eigenvalue $\alpha$.  They are related to the vacuum
state, $|0\rangle$, by a unitary transformation of the form
\begin{eqnarray}
\label{Eqn4}
|\alpha\rangle &=& \hat{D}(\alpha)|0\rangle \nonumber \\
&=& \exp\left(\alpha\hat{a}^\dagger - \alpha^*\hat{a}\right)|0\rangle \nonumber \\
&=& e^{-|\alpha|^2/2}\sum_{n=0}^\infty \frac{\alpha^n}{\sqrt{n!}}|n\rangle .
\end{eqnarray}
If our two input modes are prepared in the coherent states $|\alpha\rangle$ and $|\beta\rangle$ then we can calculate
the output state using the relationship relationship, (\ref{Eqn3}), between our input and output operators.  It is instructive
to follow the setps:
\begin{eqnarray}
\label{Eqn5}
|\alpha\rangle_{a_{\rm in}}|\beta\rangle_{b_{\rm in}} &=& \exp[\alpha\hat{a}^\dagger_{\rm in} - \alpha^*\hat{a}_{\rm in}]
\exp[\beta\hat{b}^\dagger_{\rm in} - \beta^*\hat{b}_{\rm in}]|00\rangle  \nonumber \\
&=& \exp[(t\alpha+r\beta)\hat{a}^\dagger_{\rm out} - (t^*\alpha^*+r^*\beta^*)\hat{a}_{\rm out}]  \nonumber \\
& & \qquad \times \exp[(t\beta+r\alpha)\hat{b}^\dagger_{\rm out} - (t^*\beta^*+r^*\alpha^*)\hat{b}_{\rm out}]|00\rangle  \nonumber \\
&=& |t\alpha + r\beta\rangle_{a_{\rm out}}|t\beta + r\alpha\rangle_{b_{\rm out}} ,
\end{eqnarray}
where $|00\rangle$ is the two-mode vacuum state.  We see that the complex amplitudes associated with the coherent states 
combine on the beam splitter in exactly the same way as would the field amplitudes in classical optics:
\begin{eqnarray}
\label{Eqn6}
\alpha_{\rm out} &=& t\alpha_{\rm in} + r\beta_{\rm in}  \nonumber  \\
\beta_{\rm out} &=& t\beta_{\rm in} + r\alpha_{\rm in} ,
\end{eqnarray}
so these amplitudes are related in the same way as the corresponding annihilation operators.  A simple physical way to 
understand this idea is to picture the coherent state as a superposition of a classical field with complex amplitude $\alpha$
and the vacuum state \cite{Mollow,Pegg,PLKPaul}.  The former interferes as a classical field and the latter is unaffected by the beam splitter.

\subsection{Interfering number states}

The interference at a beam splitter between modes with a well-defined photon number is more complicated than the 
interference between coherent states owing to features associated with the statistics of indistinguishable bosons.
Nevertheless, applying the beam-splitter transformation presents no special difficulties.  Let us consider the general
case in which we have $m$ photons in mode $a_{\rm in}$ and $n$ in mode $b_{\rm in}$, so that our input state is
\begin{eqnarray}
\label{Eqn7}
|\psi_{\rm in}\rangle &=& |m\rangle_{a_{\rm in}}|n\rangle_{b_{\rm in}}  \nonumber \\
&=& \frac{\left(\hat{a}^\dagger_{\rm in}\right)^m}{\sqrt{m!}}\frac{\left(\hat{b}^\dagger_{\rm in}\right)^n}{\sqrt{n!}}|00\rangle .
\end{eqnarray}
We can obtain the output state using the transformation (\ref{Eqn3}) to give
\begin{equation}
\label{Eqn8}
|\psi_{\rm out}\rangle = \frac{1}{\sqrt{m!n!}}\left(t\hat{a}^\dagger_{\rm out} + r\hat{b}^\dagger_{\rm out}\right)^m
\left(t\hat{b}^\dagger_{\rm out} + r\hat{a}^\dagger_{\rm out}\right)^n|00\rangle ,
\end{equation}
which is an entangled superposition of two-mode number states ranging from $|m+n\rangle_{a_{\rm out}}|0\rangle_{b_{\rm out}}$ 
to $|0\rangle_{a_{\rm out}}|m+n\rangle_{b_{\rm out}}$.

To demonstrate the subtle quantum features of this state it suffices to consider situations in which only two photons are
involved.  For the three possible input states, $|2\rangle_{a_{\rm in}}|0\rangle_{b_{\rm in}}$, $|0\rangle_{a_{\rm in}}|2\rangle_{b_{\rm in}}$
and $|1\rangle_{a_{\rm in}}|1\rangle_{b_{\rm in}}$ we find \cite{LoudonJOSAB}
\begin{eqnarray}
\label{Eqn9}
|2\rangle_{a_{\rm in}}|0\rangle_{b_{\rm in}} &\rightarrow & t^2|2\rangle_{a_{\rm out}}|0\rangle_{b_{\rm out}}
+ \sqrt{2}tr|1\rangle_{a_{\rm out}}|1\rangle_{b_{\rm out}} + r^2|0\rangle_{a_{\rm out}}|2\rangle_{b_{\rm out}}  \nonumber \\
|0\rangle_{a_{\rm in}}|2\rangle_{b_{\rm in}} &\rightarrow & r^2|2\rangle_{a_{\rm out}}|0\rangle_{b_{\rm out}}
+ \sqrt{2}tr|1\rangle_{a_{\rm out}}|1\rangle_{b_{\rm out}} + t^2|0\rangle_{a_{\rm out}}|2\rangle_{b_{\rm out}}  \nonumber \\
|1\rangle_{a_{\rm in}}|1\rangle_{b_{\rm in}} &\rightarrow & \sqrt{2}tr|2\rangle_{a_{\rm out}}|0\rangle_{b_{\rm out}}
+ (t^2+r^2)|1\rangle_{a_{\rm out}}|1\rangle_{b_{\rm out}} + \sqrt{2}tr|0\rangle_{a_{\rm out}}|2\rangle_{b_{\rm out}} .  \nonumber \\
& & 
\end{eqnarray}
For the first two of these we see that the probabilities for the numbers of photons in each output mode are simply 
those that we would expect if each photon were transmitted with probability $|t|^2$ and reflected with probability
$|r|^2$.  The situation is more interesting if we have one photon in each input mode.  We find two photons in output
mode $a_{\rm out}$ with probability $2|t|^2|r|^2$ or two photons in output mode $b_{\rm out}$ with the same probability.
The probability for one photon appearing in each of the output modes is $|t^2+r^2|^2 = (|t|^2 - |r|^2)^2$ which, for a 
balanced beam splitter with equal single-photon transmission and reflection coefficients is zero.  This lack of coincidences 
in the output modes is the famous Hong-Ou-Mandel effect \cite{HOM}, which has become a familiar tool in quantum optics and plays a central 
role in the proposal for optical quantum computing \cite{KLM}.

\subsection{Interfering squeezed vacuum states}

The single-mode squeezed vacuum state is generated from the vacuum by means of a two-photon analogue of the 
coherent state displacement operator \cite{Methods}:
\begin{eqnarray}
\label{Eqn10}
|\zeta\rangle &=& \hat{S}(\zeta)|0\rangle \nonumber \\
&=& \exp\left(-\frac{\zeta}{2}\hat{a}^{\dagger 2} + \frac{\zeta^*}{2}\hat{a}^2\right)|0\rangle ,
\end{eqnarray}
which is a superposition of only even photon number states 
\begin{equation}
\label{Eqn11}
|\zeta\rangle = \sqrt{{\rm sech} \: s}\sum_{n=0}^\infty \frac{\sqrt{(2n)!}}{n!}\left(-\frac{e^{i\phi}}{2}\tanh s\right)^n|2n\rangle ,
\end{equation}
where $\zeta = se^{i\phi}$ is the squeezing parameter.  

If we combine two squeezed vacuum states on a beam splitter the form of the output state will depend on the squeezing 
parameters and, in particular, their relative phase.  In general it will be a two-mode Gaussian state of Gausson 
\cite{BBBook,BBpaper,Fan,Gordon} that may 
be separable, strongly entangled or lie between these extremes.  Let us consider a simple example in which the two input
modes are prepared in squeezed vacuum states with equal squeezing strength but not necessarily squeezed in the same
quadrature, so that our input state is 
\begin{equation}
\label{Eqn12}
|\psi_{\rm in}^{\rm sq}\rangle = \hat{S}_{a_{\rm in}}(-s)\hat{S}_{b_{\rm in}}\left(-se^{i\phi}\right)|00\rangle .
\end{equation}
(We make this choice so as to arrive at a simpler description of the state in the Janszky representation.)
It is straightforward to apply the operator transformation (\ref{Eqn3}) to obtain the output state in the form
\begin{eqnarray}
\label{Eqn13}
|\psi_{\rm out}^{\rm sq}\rangle &=& \exp\left[\frac{s}{2}\left(t\hat{a}^\dagger_{\rm out} + r\hat{b}^\dagger_{\rm out}\right)^2 - H.C.\right] \nonumber \\
& & \quad \times \exp\left[\frac{se^{i\phi}}{2}\left(t\hat{b}^\dagger_{\rm out} + r\hat{a}^\dagger_{\rm out}\right)^2 - H.C.\right]|00\rangle \nonumber \\
&=& \exp\left[\frac{s}{2}\left((t^2+r^2e^{i\phi})\hat{a}_{\rm out}^{\dagger 2} + (r^2 + t^2e^{i\phi})\hat{b}_{\rm out}^{\dagger 2} \right.\right.  \nonumber \\
& & \qquad \left.\left. +2tr(1+e^{i\phi})\hat{a}_{\rm out}^\dagger\hat{b}_{\rm out}^\dagger\right) - H.C. \right]|00\rangle ,
\end{eqnarray}
where \textit{H.C.} denotes the Hermitian conjugate.  Note that for a balanced beam splitter we have $|t|^2 = \frac{1}{2} = |r|^2$ and $t^2 + r^2 = 0$
and we find the two output modes in a single-mode squeezed state for $\phi = \pi$ and in a maximally entangled two-mode squeezed 
state if $\phi = 0$ \cite{Simon}:
\begin{equation}
\label{Eqn14}
|\psi_{\rm out}^{\rm sq}(\phi = 0)\rangle = {\rm sech} \: s \sum_{n=0}^\infty \left(-\exp(i\arg tr)\tanh s\right)^n |n\rangle_{a_{\rm out}} |n\rangle_{b_{\rm out}}  .
\end{equation}

\section{The Janszky representation}
\pst

The coherent states are not mutually orthogonal but they are complete, or over-complete, in that they form a resolution of
the identity operator \cite{GlauberCoh2}:
\begin{equation}
\label{Eqn15}
\frac{1}{\pi}\int d^2\alpha |\alpha\rangle\langle\alpha| = 1
\end{equation}
where the integral runs over the whole complex $\alpha$ plane.  It follows that any state vector $|\psi\rangle$ can be expressed as an 
integral over the coherent states \cite{GlauberCoh2,Nussenzveig}:
\begin{eqnarray}
\label{Eqn16}
|\psi\rangle &=& \frac{1}{\pi}\int d^2\alpha |\alpha\rangle\langle\alpha|\psi\rangle \nonumber \\
&=& \frac{1}{\pi}\int d^2\alpha e^{-|\alpha|^2/2}\psi(\alpha^*)|\alpha\rangle ,
\end{eqnarray}
where $\psi(\alpha^*)$ is an analytic function of $\alpha^*$.  We could base our description of the field state on this function and describe 
a beam splitter in terms of it, but instead employ a different and physically appealing representation devised by Janszky and his colleagues.

Discrete superpositions of a finite number of coherent states have attracted much attention for their strongly non-classical properties
and examples include the Schr\"{o}dinger cat states, the compass states and their generalisations \cite{Haroche,Zurek,White}.  
The Janszky representation differs from 
these and the Glauber representation (\ref{Eqn16}) in that it writes any single-mode pure state as a continuous superposition of the coherent
states on a single line in the complex plane.  There is considerable freedom in how to choose the line and we consider, here, only two examples:
a circular path centred on $\alpha = 0$ and a straight line passing through $\alpha = 0$, which provide convenient representations of the number
and squeezed vacuum states respectively.

\subsection{Number states}

The number states have no preferred phase \cite{EPL88,PRA89,JMO89} and it is doubtless this property that led 
Janszky, Domokos and Adam to select a circular path centred on
the origin in the complex $\alpha$ plane \cite{JJ93,JJ94}.  If we integrate the coherent states around such a path of radius $|\alpha|$ weighted by 
$e^{-in\theta}$ then we get a state proportional to the number-state $|b\rangle$:
\begin{eqnarray}
\label{Eqn17}
\int_{0}^{2\pi} d\theta e^{-in\theta} \left||\alpha|e^{i\theta}\right\rangle &=&  e^{-|\alpha|^2/2}2\pi \frac{|\alpha|^n}{\sqrt{n!}}|n\rangle \nonumber \\
\qquad \Rightarrow \qquad  \quad |n\rangle &=& e^{|\alpha|^2/2}\frac{\sqrt{n!}}{2\pi |\alpha|^n}\int_{0}^{2\pi} d\theta e^{-in\theta} 
\left||\alpha|e^{i\theta}\right\rangle .
\end{eqnarray} 
This is the Janszky representation of the number state.  Note that we can choose any value of $|\alpha|$, corresponding to any choice
of radius for our integration path.

\subsection{Squeezed vacuum states}

The Janszky representation of the squeezed vacuum state is a Gaussian weighted integral of the coherent states along a straight line 
in the direction of the anti-squeezed quadrature.  Consider the normalised state \cite{JJ90}
\begin{equation}
\label{Eqn18}
|\psi^{\rm sq}\rangle = \pi^{-1/2}\gamma^{1/2}(\gamma^2 + 1)^{1/4}\int_{-\infty}^\infty dx e^{-\gamma^2x^2}|x\rangle ,
\end{equation}
where $|x\rangle$ is the coherent state for which $\alpha$ takes the real value $x$.  We can confirm that this superposition of coherent states
is indeed a squeezed vacuum state by evaluating its probability amplitudes in the number-state representation:
\begin{eqnarray}
\label{Eqn19}
\langle 2n+1|\psi^{\rm sq}\rangle &=& 0 \nonumber \\
\langle 2n|\psi^{\rm sq}\rangle &=& \frac{\gamma^{1/2}(\gamma^2 + 1)^{1/4}}{\left(\gamma^2 + \frac{1}{2}\right)^{1/2}}
\frac{\sqrt{(2n)!}}{n!}\left[\frac{1}{4\left(\gamma^2 + 1\right)}\right]^n .
\end{eqnarray}
This is the squeezed vacuum state (\ref{Eqn11}) with $\phi = \pi$ if we make the identification
\begin{equation}
\label{Eqn20}
\tanh s = \frac{1}{2\left(\gamma^2 + \frac{1}{2}\right)}
\end{equation}
so that 
\begin{equation}
\label{Eqn21}
e^{2s} = \frac{\gamma^2 + 1}{\gamma^2} .
\end{equation}
The variance in the in-phase quadrature $(\hat{a}+\hat{a}^\dagger)$ is increased by this factor and that in the in-quadrature quadrature
$(-i[\hat{a}-\hat{a}^\dagger])$ is squeezed by the corresponding factor, $e^{-2s}$.  It is worth pausing to examine the limiting behaviour
of this state.  For small values of $\gamma^2$ the Gaussian superposition is very broad and many coherent states participate, the 
interference of which leads to strong squeezing.  As $\gamma^2$ tends to infinity, the Gaussian in the superposition becomes vey narrow
tending, in this limit, to a delta function and centred at the origin and the state reduces to the vacuum state $|0\rangle$.

The representation of a state squeezed in a different quadrature is obtained by integrating along a different straight line path through
the origin:
\begin{equation}
\label{Eqn22}
|\zeta\rangle = \pi^{-1/2}\gamma^{1/2}(\gamma^2 + 1)^{1/4} \int_{-\infty}^\infty dx e^{-\gamma^2x^2}\left|ixe^{i\phi/2}\right\rangle  .
\end{equation}
Note that it is the angle $\phi/2$ that appears in this superposition; adding $\pi$ to $\phi$ changes a state squeezed in the imaginary 
quadrature to one that is squeezed in the real quadrature, which rotates the state through the angle $\pi/2$ in the complex $\alpha$ plane.

\section{Schr\"{o}dinger theory of the beam splitter}

The action of the beam splitter on coherent state inputs, Eq. (\ref{Eqn5}), provides a simple way to transform the input state to the output state
if we can write the initial state in terms of the coherent states.  To do this we could use Glauber's representation (\ref{Eqn16}) or Janszky's.
Here we employ the latter and, as with the conventional approach presented above, the principle is best illustrated by some examples.

\subsection{Interfering number states}

Let us return to the interference between two number states, $|m\rangle_{a_{\rm in}}|n\rangle_{b_{\rm in}}$, and write this input state in the
Janszky representation:
\begin{equation}
\label{Eqn23}
|\psi_{\rm in}^{m.n}\rangle = e^{|\alpha|^2}\frac{\sqrt{m!n!}}{4\pi^2|\alpha|^{m+n}}\int_0^{2\pi}d\theta\int_0^{2\pi}d\theta' e^{-im\theta}
e^{-in\theta'}\left||\alpha|e^{i\theta}\right\rangle_{a_{\rm in}}\left||\alpha|e^{i\theta'}\right\rangle_{b_{\rm in}}
\end{equation}
where, for simplicity, we have chosen the radii of the two integration paths to be equal.  It is straightforward to write down the
corresponding output state using the transformation law for the coherent states, Eq. ({\ref{Eqn5}).  We find
\begin{eqnarray}
\label{Eqn24}
|\psi_{\rm out}^{m.n}\rangle &=& e^{|\alpha|^2}\frac{\sqrt{m!n!}}{4\pi^2|\alpha|^{m+n}}\int_0^{2\pi}d\theta\int_0^{2\pi}d\theta' e^{-im\theta}
e^{-in\theta'}  \nonumber \\
& & \qquad \qquad \times \left||\alpha|(te^{i\theta}+re^{i\theta'})\right\rangle_{a_{\rm out}}
\left||\alpha|(te^{i\theta'}+re^{i\theta})\right\rangle_{b_{\rm out}} .
\end{eqnarray}
Obtaining from this the amplitudes for the number states at the output presents no difficulties but, for the purposes of illustration, we consider 
here only the two-photon output states considered above:
\begin{eqnarray}
\label{Eqn25}
|\psi_{\rm out}^{2,0}\rangle &=& \frac{1}{4\pi^2}\int_0^{2\pi}d\theta\int_0^{2\pi}d\theta'e^{-i2\theta}
\left[(te^{i\theta} + re^{i\theta'})^2|2\rangle_{a_{\rm out}}|0\rangle_{b_{\rm out}}  \right.  \nonumber \\
& &     \qquad + \sqrt{2}(te^{i\theta} + re^{i\theta'})(te^{i\theta'} + re^{i\theta})|1\rangle_{a_{\rm out}}|1\rangle_{b_{\rm out}} \nonumber \\
& &   \qquad \qquad \left. + (te^{i\theta'} + re^{i\theta})^2|0\rangle_{a_{\rm out}}|2\rangle_{b_{\rm out}} \right]  \nonumber \\
&=&  t^2|2\rangle_{a_{\rm out}}|0\rangle_{b_{\rm out}}
+ \sqrt{2}tr|1\rangle_{a_{\rm out}}|1\rangle_{b_{\rm out}} + r^2|0\rangle_{a_{\rm out}}|2\rangle_{b_{\rm out}}  \nonumber \\
|\psi_{\rm out}^{0,2}\rangle &=& \frac{1}{4\pi^2}\int_0^{2\pi}d\theta\int_0^{2\pi}d\theta'e^{-i2\theta'}
\left[(te^{i\theta} + re^{i\theta'})^2|2\rangle_{a_{\rm out}}|0\rangle_{b_{\rm out}}  \right.  \nonumber \\
& &     \qquad + \sqrt{2}(te^{i\theta'} + re^{i\theta})(te^{i\theta'} + re^{i\theta})|1\rangle_{a_{\rm out}}|1\rangle_{b_{\rm out}} \nonumber \\
& &   \qquad \qquad \left. + (te^{i\theta'} + re^{i\theta})^2|0\rangle_{a_{\rm out}}|2\rangle_{b_{\rm out}} \right]  \nonumber \\
&=&  r^2|2\rangle_{a_{\rm out}}|0\rangle_{b_{\rm out}}
+ \sqrt{2}tr|1\rangle_{a_{\rm out}}|1\rangle_{b_{\rm out}} + t^2|0\rangle_{a_{\rm out}}|2\rangle_{b_{\rm out}} \nonumber \\
|\psi_{\rm out}^{1,1}\rangle &=& \frac{1}{4\pi^2}\int_0^{2\pi}d\theta\int_0^{2\pi}d\theta'e^{-i(\theta+\theta')}
\left[\frac{(te^{i\theta} + re^{i\theta'})^2}{\sqrt{2}}|2\rangle_{a_{\rm out}}|0\rangle_{b_{\rm out}}  \right.  \nonumber \\
& &     \qquad + (te^{i\theta'} + re^{i\theta'})(te^{i\theta'} + re^{i\theta})|1\rangle_{a_{\rm out}}|1\rangle_{b_{\rm out}} \nonumber \\
& &   \qquad \qquad \left. + \frac{(te^{i\theta'} + re^{i\theta})^2}{\sqrt{2}}|0\rangle_{a_{\rm out}}|2\rangle_{b_{\rm out}} \right]  \nonumber \\
&=&  \sqrt{2}tr|2\rangle_{a_{\rm out}}|0\rangle_{b_{\rm out}}
+ (t^2+r^2)|1\rangle_{a_{\rm out}}|1\rangle_{b_{\rm out}} + \sqrt{2}tr|0\rangle_{a_{\rm out}}|2\rangle_{b_{\rm out}} ,
\end{eqnarray}
in agreement with the states given in Eq. (\ref{Eqn9}), calculated by the conventional approach.

\subsection{Interfering squeezed vacuum states}
\pst
For the squeezed vacuum states we employ the straight line Janszky representation described above.  For the input state in
Eq. (\ref{Eqn12}) we find
\begin{eqnarray}
\label{Eqn26}
|\psi_{\rm in}^{\rm sq}\rangle &=& \hat{S}_{a_{\rm in}}(-s)\hat{S}_{b_{\rm in}}(-se^{i\phi})|00\rangle \nonumber \\
&=& \pi^{-1}\gamma(\gamma^2 + 1)^{1/2}\int_{-\infty}^\infty dx \int_{-\infty}^\infty dx' e^{-\gamma^2(x^2+x'^2)}
|x\rangle_{a_{\rm in}}|x'e^{i\phi/2}\rangle_{b_{\rm in}} ,
\end{eqnarray}
where $\gamma^2 = (e^{2s} - 1)^{-1}$.  We obtain the output state by using, once again, the beam splitter transformation 
for the coherent states to give:
\begin{eqnarray}
\label{Eqn27}
|\psi_{\rm out}^{\rm sq}\rangle &=& \pi^{-1}\gamma(\gamma^2 + 1)^{1/2}\int_{-\infty}^\infty dx \int_{-\infty}^\infty dx' e^{-\gamma^2(x^2+x'^2)}
\nonumber \\
& & \qquad \qquad \qquad \times |tx + rx'e^{i\phi/2}\rangle_{a_{\rm in}}|tx'e^{i\phi/2}+rx\rangle_{b_{\rm in}} .
\end{eqnarray}
It is straightforward to show that this is indeed the state, Eq. (\ref{Eqn13}), calculated using the conventional approach.
To illustrate this point, we consider the two special cases discussed earlier in which the output is either two single-mode
squeezed states or a two-mode squeezed state.  Recall that these two situations occur for a balanced beam splitter, 
with $|t|^2 = \frac{1}{2} = |r|^2$ and $t^2 + r^2 = 0$, with the phase $\phi$ set to $\pi$ or $0$ respectively.  For simplicity 
of presentation we set $t = \frac{1}{\sqrt{2}}$ and $r = \frac{i}{\sqrt{2}}$.  If we set $\phi = \pi$, we find
\begin{eqnarray}
\label{Eqn28}
|\psi_{\rm out}^{\rm sq}(\phi = \pi)\rangle &=& \pi^{-1}\gamma(\gamma^2 + 1)^{1/2}\int_{-\infty}^\infty dx \int_{-\infty}^\infty dx' e^{-\gamma^2(x^2+x'^2)}
\nonumber \\
& & \qquad \qquad \qquad \times \left|\frac{1}{\sqrt{2}}(x - x')\right\rangle_{a_{\rm out}}\left|\frac{i}{\sqrt{2}}(x' + x)\right\rangle_{b_{\rm out}}
\nonumber \\
&=& \pi^{-1}\gamma(\gamma^2 + 1)^{1/2}\int_{-\infty}^\infty dx_+ \int_{-\infty}^\infty dx_- e^{-\gamma^2(x_+^2+x_-^2)} \nonumber \\
& & \qquad \qquad \qquad \times |x_-\rangle_{a_{\rm out}}|ix_+\rangle_{b_{\rm out}},  
\end{eqnarray}
where $x_\pm = (x \pm x')/\sqrt{2}$.  This is clearly a tensor product of two single-mode squeezed vacuum states with mode
$a_{\rm out}$ squeezed in the real quadrature and mode $b_{\rm out}$ squeezed in the imaginary quadrature.

The case $\phi = 0$, which should correspond to the two-mode squeezed state, requires a bit more care.  In this case
we find 
\begin{eqnarray}
\label{Eqn28}
|\psi_{\rm out}^{\rm sq}(\phi = 0)\rangle &=& \pi^{-1}\gamma(\gamma^2 + 1)^{1/2}\int_{-\infty}^\infty dx \int_{-\infty}^\infty dx' e^{-\gamma^2(x^2+x'^2)}
\nonumber \\
& & \qquad \qquad \qquad \times \left|\frac{1}{\sqrt{2}}(x + ix')\right\rangle_{a_{\rm out}}\left|\frac{i}{\sqrt{2}}(x - ix')\right\rangle_{b_{\rm out}} ,
\end{eqnarray}
which no longer resembles a pair of simple superpositions of coherent states along a line.  To proceed let us identify $x+ix'$
with the complex number $z$ and so write 
\begin{equation}
\label{Eqn30}
|\psi_{\rm out}^{\rm sq}(\phi = 0)\rangle = \pi^{-1}\gamma(\gamma^2 + 1)^{1/2} \int d^2z e^{-\gamma^2|z|^2}\left|\frac{z}{\sqrt{2}}\right\rangle 
\left|i\frac{z^*}{\sqrt{2}}\right\rangle ,
\end{equation}
where now the integral runs over the whole of the complex $z$ plane.  If we expand the two coherent states in the integrand in the number
basis and evaluate the integrals using polar coordinates then we recover the two-mode squeezed vacuum state, Eq. (\ref{Eqn14}), as we
should.

\section{Conclusion}
\pst
We have revisited the theory of the optical beam splitter making use of the coherent state representation pioneered by J\'{o}zsef Janszky.
The principal idea that makes this possible is the very simple beam-splitter transformation law for the coherent states, Eq. (\ref{Eqn5}).  The
essentially classical behaviour of the coherent states when passing through a beam splitter does not preclude the description of nonclassical
effects such as Hong-Ou-Mandel interference and the entanglement of input squeezed states.  The origin of these lies not in the coherent state
basis as such but rather in the superposition of these states.

The theory presented is general and applies to all pure-state inputs.  An extension to mixed states should be straightforward, but for these 
the phase-space quasi-probability distributions will probably provide a simpler approach.

\section*{End note}

I had the pleasure of meeting J\'{o}zsef Janszky many times over the years, starting in the early 1990s and was fortunate to visit him once 
in Budapest.  He was a genuine enthusiast for our chosen discipline and always wore a smile when we discussed physics together.  His
work was characterised by an understated elegance, typified by the Janszky representation discussed here.

\section*{Acknowledgments}
\pst
This work was supported by a Royal Society Research Professorship (RP150122).

\end{document}